\documentclass[pra,twocolumn,showpacs,amsmath,amssymb,superscriptaddress]{revtex4}

\usepackage{graphicx}
\usepackage{latexsym}
\usepackage{amsmath,amssymb}
\usepackage{verbatim,times,bbm}
\usepackage{color}

\newtheorem{definition}{Definition}

\newcommand{\Tr}{{\mathrm{Tr}}}

\begin{document}

\title{Forgetfulness of continuous Markovian quantum channels}
\author{Cosmo Lupo}
\affiliation{Dipartimento di Fisica, Universit\`{a} di Camerino,
I-62032 Camerino, Italy, EU}

\author{Laleh Memarzadeh} \affiliation{Dipartimento di Fisica,
Universit\`{a} di Camerino, I-62032 Camerino, Italy, EU}

\author{Stefano Mancini}
\affiliation{Dipartimento di Fisica, Universit\`{a} di Camerino,
I-62032 Camerino, Italy, EU} \affiliation{INFN, Sezione di Perugia,
I-06123 Perugia, Italy, EU}

\begin{abstract}

The notion of forgetfulness, used in discrete quantum memory
channels, is slightly weakened in order to be applied to the case of
continuous channels. This is done in the context of quantum memory
channels with Markovian noise. As a case study, we apply the notion
of weak-forgetfulness to a bosonic memory channel with additive
noise. A suitable encoding and decoding unitary transformation
allows us to unravel the effects of the memory, hence the channel
capacities can be computed using known results from the memoryless
setting.

\end{abstract}

\pacs{03.67.Hk, 05.40.Ca, 42.50.-p, 89.70.-a}

\maketitle

\section{Introduction}

One of the main issues in quantum information theory is the
evaluation of the maximum rate, i.e.\ the capacity, at which
(classical or quantum) information can be reliably transmitted via a
quantum communication channel. When studying models of noisy quantum
communication, a common assumption is that the noise affecting the
channel is identical and independent at each channel use. In
mathematical terms, the completely positive trace-preserving (CPT)
map $\mathcal{E}^{(n)}$ describing $n$ uses of the quantum channel
is the direct product of $n$ identical copies:
\begin{equation}\label{memory-less}
\mathcal{E}^{(n)} = \bigotimes_{j=1}^n \mathcal{E},
\end{equation}
where $\mathcal{E}$ is the CPT map describing a single use of the
quantum channel. A channel of this kind is called, as its classical
counterpart, a memoryless quantum channel. Coding theorems for
memoryless quantum channels, allowing to write the channel
capacities in terms of entropic quantities, are well established
results in quantum information theory \cite{coding}. However, the
assumption of independent and identical noise can be rather
artificial in several physical settings where memory effects may
naturally appear, see e.g.\ \cite{KW2} and references therein. This
observation leads to consider quantum channels with a more general
structure than the simple tensor-product structure of
(\ref{memory-less}). Every quantum channel such that
\begin{equation}
\mathcal{E}^{(n)} \neq \bigotimes_{j=1}^n \mathcal{E}
\end{equation}
is called a quantum channel with memory, or simply a {\it memory
channel}. For memory channels the noises affecting multiple channel
uses are in general neither independent nor identical.

\begin{figure}
\centering
\includegraphics[width=0.4\textwidth]{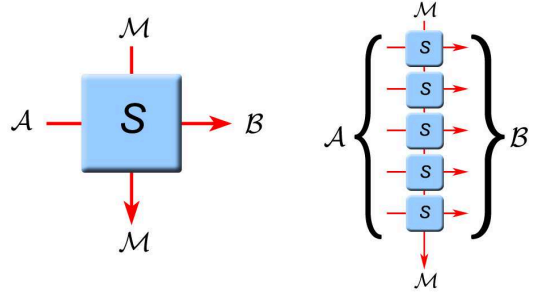}
\caption{On the left: each use of the memory channel is represented
by an elementary transformation $S$ with two input systems
$\mathcal{A}$ and $\mathcal{M}$ and two outputs $\mathcal{B}$ and
$\mathcal{M}$. On the right: $n$ uses of the memory channel are
represented as the $n$-fold concatenation of the elementary
transformation.} \label{channel}
\end{figure}

The structure theorem for memory channels was provided in
\cite{KW2}. Under the assumptions of causality and invariance under
time translation, a sequence of $n$ uses of a memory channel can be
always decomposed as the $n$-fold concatenation $\mathsf{S}^{(n)}$
of an elementary transformation $\mathsf{S}$. Such decomposition
requires the introduction of an ancillary system $\mathcal{M}$,
called the {\it memory kernel} (or simply the {\it memory}), which
accounts for correlations. Such elementary transformation has two
input and two output systems. In Fig.~\ref{channel} the horizontal
lines indicates the sender ($\mathcal{A}$) and the receiver
($\mathcal{B}$) systems, the vertical line the input and output
memory. Multiple uses of the memory channel are hence obtained by
concatenating the elementary transformation through the vertical
line, as shown in the right hand side of Fig.~\ref{channel}.

The performances of the memory channel are in general determined by
the memory initialization. Different initial states of the memory
kernel can lead to different values of the channel capacities. This
is not the case for ``forgetful'' channels, whose capacities are
independent on the memory initialization. Moreover, coding theorems
for forgetful channels are straightforward extensions of their
memoryless counterparts. The behavior of forgetful channels is
asymptotically independent on the memory initialization, hence the
memory system, after a sufficiently large number of channel uses,
``forgets'' what was its initial state. This property was put
forward in \cite{Bowen}. Then, the notion of ``forgetfulness' in
discrete quantum channels, i.e.\ CPT map acting on finite
dimensional Hilbert spaces, has been formalized in \cite{KW2}.

Recently, in the framework of continuous memory channel, i.e.\ CPT
map acting on infinite dimensional Hilbert spaces, it has been
noticed that the extension of the notion of forgetfulness to this
framework is highly nontrivial \cite{LGM}.

Below, this notion will be slightly weakened and extended to the
case of continuous quantum channels by considering Markovian noise.
As an application we shall evaluate the classical capacity of a
bosonic memory channel with additive noise.

The paper develops along the following lines. In Sec.~\ref{Markov}
the notion of forgetfulness will be considered in the context of
quantum channels with Markovian correlated noise and adapted to the
continuous variable setting. We introduce a notion of
``weak-forgetfulness'' to be applied in the case of a Markov process
with continuous noise variable. In Sec.~\ref{model} a model of
quantum channel subjected to additive Gaussian noise with Markovian
correlations will be proposed. Suitable encoding and decoding
unitary transformations allow us to unravel the effects of the
memory, hence the channel capacities can be computed using known
results from the memoryless setting.

\section{Quantum channels with Markovian correlated noise}\label{Markov}

In this section we consider the notion of forgetfulness applied to
the case of a class of quantum memory channels with Markovian
correlated noise.

Let us first recall the definition of forgetfulness as presented in
\cite{KW2}.

\begin{definition}[Forgetfulness] A memory channel is forgetful iff for any
$\epsilon > 0$, there exists an integer $\nu$ such that for any $n
> \nu$
\begin{equation}\label{def1}
\left\|
\Tr_\mathcal{B}\left[\mathsf{S}^{(n)}(\rho_{1,\mathcal{A},\mathcal{M}})\right]
-
\Tr_\mathcal{B}\left[\mathsf{S}^{(n)}(\rho_{2,\mathcal{A},\mathcal{M}})\right]
\right\|_1 < \epsilon \, ,
\end{equation}
for any $\rho_{1,\mathcal{A},\mathcal{M}}$,
$\rho_{2,\mathcal{A},\mathcal{M}}$ states of the $n$ inputs and the
initial memory such that
\begin{equation}\label{mem_init}
\Tr_\mathcal{M}(\rho_{1,\mathcal{A},\mathcal{M}})=\Tr_\mathcal{M}(\rho_{2,\mathcal{A},\mathcal{M}})
\, .
\end{equation}
\end{definition}

This definition of forgetfulness applies in the Schroedinger picture
description of the memory channel, an equivalent definition can be
formulated in the Heisenberg picture. Let us briefly comment it. The
density operators $\rho_{1,\mathcal{A},\mathcal{M}}$,
$\rho_{2,\mathcal{A},\mathcal{M}}$ describe two input states of the
$n$-fold concatenation $\mathsf{S}^{(n)}$, including the initial
state of the memory kernel $\mathcal{M}$ and the state on the $n$
channel inputs belonging to the sender $\mathcal{A}$. Equation
(\ref{mem_init}) states that $\rho_{1,\mathcal{A},\mathcal{M}}$ and
$\rho_{2,\mathcal{A},\mathcal{M}}$ only differ for the reduced state
of memory kernel, corresponding to two different memory
initializations. In Eq.~(\ref{def1}), the partial traces
$\Tr_\mathcal{B}\left[\mathsf{S}^{(n)}(\rho_{1,\mathcal{A},\mathcal{M}})\right]$,
$\Tr_\mathcal{B}\left[\mathsf{S}^{(n)}(\rho_{2,\mathcal{A},\mathcal{M}})\right]$,
over the $n$ output of the channel belonging to the receiver
$\mathcal{B}$, indicate the final states of the memory kernel after
$n$ channel uses. Hence, after $n > \nu$ uses of a forgetful channel
the final state of the memory kernel can be assumed to be
independent on the memory initialization with an error smaller than
$\epsilon$, where $\nu$ is only determined by the error threshold
$\epsilon$, uniformly for all initial states of the memory kernel.
The trace distance
\begin{equation}
\left\| \rho_1 - \rho_2 \right\|_1 := \Tr\left( \left| \rho_1 -
\rho_2 \right| \right) \, .
\end{equation}
is used to quantify the distance between the final states of the
memory kernel.

If the memory channel is forgetful, one can adopt a double-block
encoding. Over $m=n+l$ channel uses, the first $n>\nu$ are not used
to send information, but only to let the memory kernel forget its
initial state with an error smaller than $\epsilon$, then the
remaining $l$ are used to send information to the channel. This
double blocking procedure allows to prove the coding theorem for
forgetful channels.

Quantum channels with Markovian correlated noise were first
considered in \cite{BM}. Here we are going to consider such channels
characterized by a memory system represented by a classical random
variable $Z$ taking values $z \in \Omega$ in a measurable set
$\Omega$.

At the $k$th use of the channel an input state $\rho_\mathcal{A}$
maps to an output state
\begin{equation}
\rho_\mathcal{B} = \int dz P_k(z) \mathcal{E}_{z}(\rho_\mathcal{A})
\, ,
\end{equation}
where $P_k(z)$ is the probability distribution of random variable
$Z$ at step $k$ and $\mathcal{E}_{z}$ is a CPT map for any
$z\in\Omega$. The probability distribution of the noise variable
changes according to the Markov rule
\begin{equation}\label{m-rule}
P_{k+1}(z)= \int dz' w(z|z')P_k(z').
\end{equation}
in which $\omega(z|z')$ is the transition function determining the
stationary Markov process. The model studied in \cite{qubits}
belongs to this class of memory channels. Coding theorems for this
class of memory channels were provided in \cite{Datta} in the case
$z$ is a discrete variable.

Recalling that in the forgetful channel, the final state of the
memory is independent of the initial memory, one would say that a
quantum memory channel with Markovian correlated noise is forgetful
if and only if the Markov process of the environment has a unique
stationary state. This is indeed the case for quantum channels
acting on discrete variable quantum systems. From this intuition we
are led to introduce the notion of ``weak-forgetfulness'' to include
the case of infinite of infinite dimensional quantum memory
channels. Considering this we define weak-forgetful channels as
follows:

\begin{definition}[Weak-Forgetfulness] A memory channel with Markovian correlated
noise is ``weak-forgetful'' iff for any $\epsilon>0$, and for any
pair of initial probability distributions $P_1(z)$, $P'_1(z)$, there
exists an integer $\nu$ such that for any $n > \nu$
\begin{equation}
\left\| P_n - P'_n \right\| < \epsilon \,.
\end{equation}
where
\begin{equation}
\left\| P_n - P'_n \right\|=\int dz \left|P_{n}(z)-P'_n(z)\right|,
\end{equation}
is the distance between the probability distributions at step $n$
with two different initial probability distributions $P_1(z)$ and
$P'_{1}(z)$.
\begin{align}
P_n(z) & = \int \prod_{k=1}^{n-1} dz_k  w(z|z_{n-1})\cdots w(z_2|z_1)P_1(z_1)\, , \label{n-fold} \\
P'_n(z) & = \int \prod_{k=1}^{n-1} dz_k w(z|z_{n-1})\cdots
w(z_2|z_1)P'_1(z_1) \, .
\end{align}
\end{definition}

Hence we can say that, even in the case of continuous variables, a
memory channel with Markovian correlated noise is weak-forgetful iff
the underlying Markov process has unique stationary state.

To adopt a double block procedure one should wait for $n > \nu$
channel uses in order to let the noise process approaches the
stationary state and then start encoding information. It is worth to
mention that ``weak-forgetfulness'' differs from ``forgetfulness''
property in the sense that the noise probability distributions
converge not uniformly with respect to the initial distributions
$P_1$, $P'_1$. Some examples will be discussed in the next section.
In conclusion, the notion of forgetfulness and weak-forgetfulness
clearly coincide if the set $\Omega$ in which the noise variable
takes values is compact. This is the case of discrete random
variable studied in \cite{Datta}.

\section{Additive Gaussian noise}\label{model}

In this section we consider a model of bosonic memory channel with
Markovian correlated noise. The notion of weak-forgetfulness is
applied to this model.

The model under consideration is a bosonic channel with additive
noise. A sequence of $n$ uses of the memory channel maps $n$ input
bosonic modes, with ladder operators $\{ a_k , a_k^\dag
\}_{k=1,\dots n}$, onto $n$ output modes, described by the operators
$\{ b_k, b_k^\dag \}_{k=1,\dots n}$.

In the Heisenberg picture, the mode operators are transformed as
follows:
\begin{equation}
b_k = a_k + z_k \, , \quad b_k^\dag = a_k^\dag + z_k^* \, ,
\end{equation}
where $z_k \in \mathbb{C}$ is the value of the random variable $Z$
at the $k$th step.

In the Schroedinger picture, a density operator
$\rho^{(n)}_\mathcal{A}$ describing the state of the $n$ input modes
is subjected to a {\it random displacement}, i.e.\
\begin{align}\label{SPmap}
\rho^{(n)}_\mathcal{B} = \int & \left[\prod_{k=1}^n d^2 z_k\right]
P(z_1,z_2,\dots z_n) \times \nonumber\\
& \times \left[ \bigotimes_{k=1}^n \mathcal{D}_k(z_k)\right]
\rho^{(n)}_\mathcal{A} \left[\bigotimes_{k=1}^n
\mathcal{D}_k(z_k)^\dag\right],
\end{align}
where $\mathcal{D}_k(z_k)$ is the displacement operator acting on
the $k$th input mode, and $P(z_1,z_2,\dots z_n)$ is the joint
probability distribution of the $n$ noise variables.

The quantum channel is Gaussian if and only if the probability
distribution of the noise is Gaussian.

%---

Our aim is to compute the capacity of the quantum channel. In order
to avoid unphysical results, we impose a constraint on the maximum
energy at the input modes by the following condition:
\begin{equation}
\frac{1}{n} \Tr \left( \rho^{(n)}_\mathcal{A} \sum_{k=1}^n a_k^\dag
a_k \right) \le N \, .
\end{equation}

The memoryless limit is recovered iff the noise variables are
mutually independent and identically distributed, i.e.\ if and only
if the joint probability distribution is the product of $n$
identical distributions:
\begin{equation}\label{memoryless}
P(z_1,z_2,\dots z_n) = \bigotimes_{k=1}^n P(z_k) \, .
\end{equation}

%---

A remarkable case is obtained if the noise variables come from a
time-independent Markov process. In this case the quantum channel
satisfies the conditions of causality and invariance under time
translations and the structure theorem can be applied. The joint
probability distribution reads
\begin{equation}
P(z_1,z_2,\dots z_n) =  \omega(z_n|z_{n-1}) \dots \omega(z_2|z_1)
P_1(z_1) \, ,
\end{equation}
where $\omega(z_k|z_{k-1})$ is the transition function determining
the Markov chain and $P_1(z_1)$ is the initial probability
distribution describing the noise variable at the first channel use.

%---

In order to construct a Gaussian channel, one has to consider a
Gaussian stochastic process. Here we consider a Gaussian transition
function of the form:
\begin{equation}\label{trans}
\omega(z_k|z_{k-1}) \simeq \exp{\left[ - \frac{|z_k - \mu
z_{k-1}|^2}{(1-\mu^2)\sigma}\right]} \, .
\end{equation}
Here and in the following we omit writing the normalization factor
in front of the probability density distributions.

The memory channel is hence described by two parameters. The
parameter $\mu \in [0,1]$ accounts for the memory effects, and
$\sigma \ge 0$, as it will be made clear below, to the amount of
noise in the channel. The memoryless limit is recovered for $\mu=0$,
in which case the joint probability distribution factorizes as in
(\ref{memoryless}).

%---

The features of the memory channel depends on the underlying Markov
process. Inserting (\ref{trans}) into (\ref{n-fold}) we obtain
\begin{equation}
P_n(z) \simeq  \int dz_1 \exp{\left( - \frac{|z - \mu^n
z_1|^2}{\sigma_n} \right)} P_1(z_1) \, ,
\end{equation}
where $\sigma_n = \sigma (1-\mu^{2n})$. By considering the limit
$n\to\infty$ we distinguish the following cases.

\subsection{Noise process at the stationary state}

For $\mu \in [0,1[$ and $\sigma>0$ there exists an unique stationary
distribution
\begin{equation}
P_s(z) \simeq \exp{\left( - \frac{z^* z}{\sigma} \right)} \, .
\end{equation}
Hence, for these values of the parameters, the memory channel is
weak-forgetful. Notice that the parameter $\sigma$ is the noise
variance of the stationary distribution.

Considering the stationary state of the Markov process is hence
sufficient for computing the channel capacities. Upon $n$ channel
uses the stationary process is described by a Gaussian joint
probability density distribution
\begin{equation}\label{case_1}
P(z_1, \dots z_n) \simeq \exp{\left[ - \frac{\sum_{hk} z_h^* M_{hk}
z_k}{(1-\mu^2)\sigma}\right]} \, ,
\end{equation}
where $M$ is the $n \times n$ tridiagonal matrix:
\begin{eqnarray}
M = \left(\begin{array}{cccccc}
1      & -\mu    & 0       & \dots  & 0       & 0 \\
-\mu   & 1+\mu^2 & -\mu    & \dots  & 0       & 0 \\
0      & -\mu    & 1+\mu^2 & -\mu   & \dots   & 0 \\
\vdots & \ddots  & \ddots  & \ddots & \ddots  & \vdots \\
0      & \dots   & 0       & -\mu   & 1+\mu^2 & -\mu \\
0      & \dots   & 0       & 0      & -\mu    & 1
\end{array}\right).
\end{eqnarray}

For any $n$, the quadratic form appearing in (\ref{case_1}) can be
always put in a diagonal form
\begin{equation}\label{diagonal}
\sum_{hk} z_h^* M_{hk} z_k = \sum_j m_j | \tilde z_j |^2 \, ,
\end{equation}
in terms of the {\it collective} noise variables
\begin{equation}
\tilde z_j := \sum_{k} O_{jk} z_k \, , \quad \tilde z_j^* :=
\sum_{k} O_{jk} z_k^* \, ,
\end{equation}
where $O$ is the $n \times n$ orthogonal matrix diagonalizing $M$:
\begin{equation}
O_{jh} M_{hk} O_{j'k} = \delta_{jj'} m_j \, .
\end{equation}

By applying a unitary encoding and decoding transformations, we can
analogously define the {\it collective} input variables
\begin{equation}\label{encoding}
\tilde a_j := \sum_{k} O_{jk} a_k \, , \quad \tilde a_j^\dag :=
\sum_{k} O_{jk} a_k^\dag \, ,
\end{equation}
and output variables
\begin{equation}
\tilde b_j := \sum_{k} O_{jk} b_k \, , \quad \tilde b_j^\dag :=
\sum_{k} O_{jk} b_k^\dag \, ,
\end{equation}
which transform according to
\begin{equation}
\tilde b_j = \tilde a_j + \tilde z_j \, , \quad \tilde b_j^\dag =
\tilde a_j^\dag + \tilde z_j^* \, .
\end{equation}

Hence, $n$ uses of the memory channel are unitary equivalent to the
tensor product of $n$ additive noise channels, whose noise variables
are mutually independent but not identically distributed. From
(\ref{diagonal}), the collective noise variables are distributed
according to the Gaussian distributions
\begin{equation}
\tilde P_j(\tilde z_j) \simeq \exp{\left( - \frac{|\tilde
z_j|^2}{\tilde \sigma_j} \right)} \, ,
\end{equation}
where the noise variances are
\begin{equation}
\tilde \sigma_j = \frac{(1-\mu^2)\sigma}{m_j} \, .
\end{equation}
For any $n$, the distribution of the noise variances can be computed
from the eigenvalues of the matrix $M$. Notice that the energy
constrain is preserved in terms of the collective input variables,
i.e.\
\begin{equation}
\frac{1}{n} \Tr \left( \rho^{(n)}_\mathcal{A} \sum_{j=1}^n \tilde
a_j^\dag \tilde a_j \right) \le N \, .
\end{equation}

In the limit of $n\to\infty$, the distribution of the eigenvalues of
the matrix $M$, arranged in nondecreasing order, tends to an
asymptotic distribution, described by the function
\begin{equation}
m^\infty(\lambda)= | 1 - \mu e^{i\lambda}|^2 \, .
\end{equation}
for $\lambda \in [0,\pi]$, in the sense that \cite{Gray}:
\begin{equation}
\lim_{n\to\infty} \frac{1}{n} \sum_j \left| m_j - m^\infty(\pi j/n)
\right| = 0 \, .
\end{equation}
Analogously, for $m_j, m^\infty(\lambda) > 0$, we have
\begin{equation}\label{as_equivalence}
\lim_{n\to\infty} \frac{1}{n} \sum_j \left| \tilde \sigma_j -
\sigma(\pi j/n) \right| = 0 \, ,
\end{equation}
where the asymptotic distribution of the noise variances, arranged
in nonincreasing order, is
\begin{equation}
\sigma(\lambda) = \sigma \frac{1-\mu^2}{| 1 - \mu e^{i\lambda}|^2}
\, .
\end{equation}

As consequence of (\ref{as_equivalence}), for any smooth function
$F$, the following equality holds true
\begin{equation}\label{average}
\lim_{n\to\infty} \frac{1}{n} \sum_{j} F(\tilde\sigma_j) =
\int_0^\pi \frac{d\lambda}{\pi} F(\sigma(\lambda)) \,.
\end{equation}

%---

\subsubsection*{Classical capacity}

The additive noise channel has been widely studied in the
memoryless, Gaussian case. We recall the case of the memoryless
broadband channel. At each use of the channel, $J$ input modes $\{
a_\kappa , a^\dag_\kappa \}_{\kappa=1,\dots J}$ are subject to
independent, but not identically distributed, Gaussian additive
noises with variances $\sigma_\kappa$. A lower bound on the
classical capacity can be obtained optimizing over Gaussian
encoding. Moreover, using the recently proven {\it minimum output
entropy conjecture}, it is possible to show that the classical
capacity of the broadband channel, per mode and expressed in bits,
is
\begin{equation}\label{broadband}
C = \frac{1}{J}\sum_\kappa g(N_\kappa + \sigma_\kappa) -
g(\sigma_\kappa) \, ,
\end{equation}
where $g(x) := (x+1) \log_2{(x+1)} - x \log_2{(x)}$ and
\begin{equation}\label{optimal}
N_\kappa = \left( \frac{1}{2^L - 1} - \sigma_\kappa \right)_+
\end{equation}
where $(x)_+$ equals $x$ if $x>0$ and is zero otherwise. The value
of the Lagrange multiplier $L$ is the root of the integral equation
\begin{equation}\label{constraint}
\frac{1}{J}\sum_{\kappa=1}^J \left( \frac{1}{2^L - 1} -
\sigma_\kappa \right)_+ = N \, .
\end{equation}

%---

Using the result for the memoryless broadband channel we can now
compute the classical capacity of the memory channel, in the region
$\mu \in [0,1[$, $\sigma > 0$, by following the same line of
reasoning of \cite{LGM}.

For any $n$, we can group the set of collective modes in $J$ blocks
of length $\ell = n/J$. At the boundaries of the $\kappa$th block
the maximum and minimum limits of the effective noise variances are
\begin{equation}
\overline \sigma_\kappa := \limsup_{n\to\infty} \tilde
\sigma_{(\kappa-1)n/J+1} \, , \quad \underline \sigma_\kappa :=
\liminf_{n\to\infty} \tilde \sigma_{\kappa n/J} \, .
\end{equation}
Recalling that the noise variances $\tilde\sigma_j$ are in
nonincreasing order, it follows that for arbitrary $\delta
> 0$ and for sufficiently large $\ell$:
\begin{equation}
\underline \sigma_\kappa - \delta \le \tilde \sigma_{(\kappa-1)\ell
+j} \le \overline \sigma_\kappa + \delta
\end{equation}
for any $\kappa$ and $j=1,\dots\ell$.

From the last equation it follows that the classical capacity of the
memory channel is bounded from above and from below by the capacity
of two memoryless broadband channels, respectively characterized by
the set of $J$ noise variances $\{ \underline \sigma_\kappa - \delta
\}_{\kappa=1,\dots J}$ and $\{ \overline \sigma_\kappa + \delta
\}_{\kappa=1,\dots J}$.

Now, keeping $J$ fixed and in the limit $\ell\to\infty$, we can
write the following bounds for the classical capacity:
\begin{equation}
\underline C_J \le C \le \overline C_J \, ,
\end{equation}
where
\begin{eqnarray}
\underline C_J & := & \frac{1}{J} \sum_{\kappa=1}^J g(\underline
N_\kappa + \overline \sigma_\kappa) -
g(\overline \sigma_\kappa) \, , \\
\overline C_J & := & \frac{1}{J} \sum_{\kappa=1}^J g(\overline
N_\kappa + \underline \sigma_\kappa) - g(\underline \sigma_\kappa)
\, ,
\end{eqnarray}
and the optimal distribution $\underline N_\kappa$, $\overline
N_\kappa$ are as in Eq.s (\ref{optimal}), (\ref{constraint}).

Finally, in the limit $J\to\infty$ the lower and upper bound
coincide. Using (\ref{average}) that yields the following formula
for the classical capacity:
\begin{equation}\label{capacity_full}
C = \int_0^\pi \frac{dz}{\pi} g[N(z)+\sigma(z)] - g[\sigma(z)] \, ,
\end{equation}
where the function $N(z)$ is determined according to the continuous
limit of Eq.s (\ref{optimal}), (\ref{constraint}), i.e.\
\begin{eqnarray}
N(z) & = & \left( \frac{1}{2^L - 1} - \sigma(z) \right)_+ \, , \label{N_full}\\
N & = & \int_0^\pi \frac{dz}{\pi} \left( \frac{1}{2^L - 1} -
\sigma(z) \right)_+ \, . \label{L_full}
\end{eqnarray}

\begin{figure}
\centering
\includegraphics[width=0.4\textwidth]{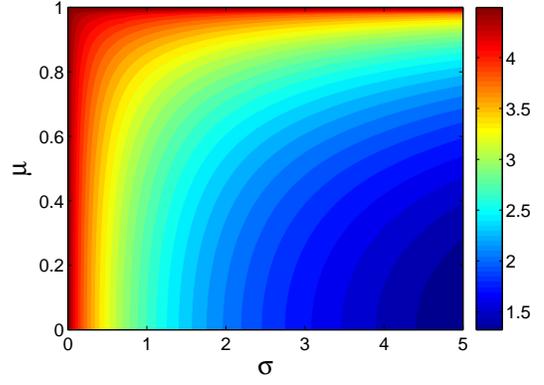}
\caption{(Color online.) The density plot shows the classical
capacity of the Markovian correlated additive noise channel as
function of the parameters $\sigma$ and $\mu$. The maximum value of
the number of excitation per mode is $N=8$, corresponding to a
noiseless channel classical capacity $g(N)\simeq 4.5293$.}
\label{classical}
\end{figure}

The formulas (\ref{capacity_full}), (\ref{N_full}), (\ref{L_full})
can be used to numerically compute the classical capacity of the
memory channel in the region $\mu \in [0,1[$ and $\sigma>0$. The
numerical results are plotted in Fig.~\ref{classical}. We remark
that, although we have assumed the noise process to be at the
stationary state, since the memory channel is weak-forgetful, the
obtained result is the classical capacity for all the initial states
of the memory.

\subsection{Critical behavior}

Some care is needed in dealing with the parameter regions defined by
$\sigma = 0$ and $\mu \in [0,1[$, and defined by $\mu = 1$. For
these values of the parameters the transition functions become
singular.
\begin{eqnarray}
\lim_{\sigma\to 0, \mu<1} \omega(z_k | z_{k-1}) & = & \delta(z_k - \mu z_{k-1}) \, , \\
\lim_{\mu\to 1} \omega(z_k | z_{k-1}) & = & \delta(z_k - z_{k-1}) \,
. \label{perfect}
\end{eqnarray}

In the limit $\sigma\to 0$, an unique stationary state exists
although singular, i.e.\ $P_s(z)=\delta(z)$. We can still say that
the channel is weak-forgetful. By noticing that the stationary state
of the Markov process corresponds to a noiseless channel, we can say
that for $\sigma=0$ the classical capacity of the memory channel is
given by the noiseless channel formula $C=g(N)$.

In the limit $\mu\to 1$, the Dirac $\delta$-function in
(\ref{perfect}) implies that the noise acting at different channel
uses are perfectly correlated. It is immediate to recognize that in
this case the Markov process has infinitely many stationary states.
The channel has hence long-term memory and is not weak-forgetful.
Thus we cannot say {\it a priori} that the channel capacity is
independent on the memory initialization. However, we can still
solve the channel by proceeding as follows. Upon $n$ channel uses
the corresponding joint probability distribution of the noise
variables reads as follows
\begin{equation}\label{ltm}
P(z_1,\dots z_n) = \delta(z_n-z_{n-1}) \cdots \delta(z_2 - z_1)
P_1(z_1) \, ,
\end{equation}
where $P_1(z_1)$ is the initial noise distribution. For a generic
initial noise distribution, even a nonGaussian one, we can solve the
problem of the channel capacity by introducing suitable
encoding/decoding unitary transformations which allow to unravel the
memory. For any $n$, we can define the collective noise variable
\begin{equation}
\tilde z_1 := \frac{1}{\sqrt{n}} \sum_{k=1}^n z_k \, ,
\end{equation}
together with a set of $n-1$ variables
\begin{equation}
\tilde z_j := \frac{1}{\sqrt{n}} \sum_{k=1}^n \exp{\left(\iota 2\pi
\frac{j-1}{n}k\right)} \, z_k \, , \, \mbox{for} \, j=2,\dots n \, .
\end{equation}
In terms of these collective noise variables, the joint probability
distribution (\ref{ltm}) factorizes as follows:
\begin{equation}
\tilde P (\tilde z_1 , \dots \tilde z_{n-1}) = P_1(\tilde z_1)
\delta(\tilde z_2) \dots \delta(\tilde z_n) \, .
\end{equation}
Hence, introducing the collective input and output variables
\begin{eqnarray}
\tilde a_1 := \frac{1}{\sqrt{n}} \sum_{k=1}^n a_k \, , \quad \tilde
a_j := \frac{1}{\sqrt{n}} \sum_{k=1}^n \exp{\left(\iota 2\pi
\frac{j-1}{n}k\right)} \, a_k \, , \nonumber\\
\tilde b_1 := \frac{1}{\sqrt{n}} \sum_{k=1}^n b_k \, , \quad \tilde
b_j := \frac{1}{\sqrt{n}} \sum_{k=1}^n \exp{\left(\iota 2\pi
\frac{j-1}{n}k\right)} \, b_k \, ,\nonumber
\end{eqnarray}
it follows that the collective mode $\{ \tilde a_1 , \tilde a_1^\dag
\}$ is subject to the additive noise described by the initial noise
probability $P_1(\tilde z_1)$, while the remaining $n-1$ collective
modes experience a noiseless channel. In conclusion, taking the
limit $n\to\infty$ and independently of the initial noise
distribution, the classical capacity of the memory channel is given
by the noiseless formula $C = g(N)$.

It is worth noticing that the classical capacity at the singular
region coincides with the analytical continuation of the expression
in Eq.~(\ref{capacity_full}).

To conclude this section we notice that the stationary Gaussian
process discussed in the previous subsection can be mapped into the
'Gaussian model' discussed in \cite{BerlinKac}. In this mapping, the
point $\mu = 1$, which gives rise to a channel with long-term
memory, corresponds to the critical point of the Gaussian model.

\section{Conclusions}

In conclusion we have considered the notion of forgetfulness for
memory channels with Markovian correlated noise. For the case of a
Markov process with discrete noise variables forgetfulness is
equivalent to the existence of unique stationary noise distribution.
In the case of continuous variable Markov process, we have
introduced a notion of weak-forgetfulness. A memory channel with
continuous variable Markovian correlated noise is weak-forgetful iff
the noise process has unique stationary distribution. Moreover the
capacities are independent of the memory initialization. The notion
of forgetfulness and weak-forgetfulness are equivalent in the
discrete variables setting. As an application, we have proposed a
model of bosonic Gaussian channel with additive Markovian correlated
noise and computed the classical capacity. The channel is either
weak-forgetful or has long-term memory. In all the cases the
classical capacity has been be computed exactly
(Fig.~\ref{classical} summarizes the obtained results).

It is worth noticing that the capacity is reached without the use of
entangled codewords. This can be easily proven by noticing that the
encoding transformation in Eq.~(\ref{encoding}) transforms coherent
states into coherent states, and recalling that coherent state
encoding is optimal to reach the memoryless classical capacity in
Eq.~(\ref{broadband}). This is related to the fact that the
considered channel model is covariant under gauge transformations
$a_k \to e^{\iota \phi} a_k$. Entangled codewords would be necessary
if one consider a noise process which breaks this symmetry, see
e.g.\ \cite{continuous}.

Other kinds of capacities can be computed along the same line of
reasoning for the considered model. Furthermore, by exploiting the
recently proven {\it minimum output entropy conjecture}
\cite{Lloyd09}, the same methods can be applied to determine the
capacities of other bosonic channels, e.g.\ attenuation and
amplification channels, with Markovian noise.

%%%

\acknowledgments The authors would like to thank V.~Giovannetti for
valuable comments. C.L.\ and L.M.\ are grateful to J.~G\"{u}tschow,
D.~Gross, and R.~F.~Werner for the stimulating discussions. The
research leading to these results has received funding from the
European Commission's seventh Framework Programme (FP7/2007-2013)
under grant agreement no.~213681. After completing the paper, we
became aware of a related work on Markovian memory channels
\cite{Cerf}.

\end{document}